\documentstyle{article}
\makeatletter

\def\@aabuffer{}
\def\author #1{\expandafter\def\expandafter\@aabuffer\expandafter
{\@aabuffer \small\rm      #1\relax \par}}
\def\address#1{\expandafter\def\expandafter\@aabuffer\expandafter
{\@aabuffer \small\it #1\relax \par\vspace{1em}}}

\def\maketitle{
\begin{center}
   {\bf \@title \par}       
   \vskip 2em                      
   \@aabuffer\relax
\end{center} \par
\gdef\@aabuffer{}
}

\def\abstracts#1{
\begin{center}
{\begin{minipage}{4.2truein}
                 \footnotesize
                 \parindent=0pt #1\par
                 \end{minipage}}\end{center}
                 \vskip 2em \par}

\fussy
\def\section{\@startsection {section}{1}{\z@}{-3.5ex plus -1ex minus 
    -.2ex}{2.3ex plus .2ex}{\bf }}
\def\subsection{\@startsection{subsection}{2}{\z@}{-3.25ex plus -1ex minus 
   -.2ex}{1.5ex plus .2ex}{\it }}

\def\@makefnmark{{$\!^{\@thefnmark}$}}

\renewenvironment{thebibliography}[1]
        {\begin{list}{\arabic{enumi}.}
        {\usecounter{enumi}\setlength{\parsep}{0pt}
         \setlength{\itemsep}{0pt} 
         \settowidth
        {\labelwidth}{#1.}\sloppy}}{\end{list}}

\newcounter{arabiclistc}

\def\@citex[#1]#2{\if@filesw\immediate\write\@auxout
        {\string\citation{#2}}\fi
\def\@citea{}\@cite{\@for\@citeb:=#2\do
        {\@citea\def\@citea{,}\@ifundefined
        {b@\@citeb}{{\bf ?}\@warning
        {Citation `\@citeb' on page \thepage \space undefined}}
        {\csname b@\@citeb\endcsname}}}{#1}}

\newif\if@cghi
\def\cite{\@cghitrue\@ifnextchar [{\@tempswatrue
        \@citex}{\@tempswafalse\@citex[]}}
\def\citelow{\@cghifalse\@ifnextchar [{\@tempswatrue
        \@citex}{\@tempswafalse\@citex[]}}
\def\@cite#1#2{{$\!^{#1}$\if@tempswa\typeout
        {IJCGA warning: optional citation argument 
        ignored: `#2'} \fi}}

\def\baselinestretch{1.0}
\ifx\selectfont\undefined
\@normalsize\else\let\glb@currsize=\relax\selectfont
\fi

\ifx\selectfont\undefined
\def\@singlespacing{%
\def\baselinestretch{1}\ifx\@currsize\normalsize\@normalsize\else\@currsize\fi%
}
\else
\def\@singlespacing{\def\baselinestretch{1}\let\glb@currsize=\relax\selectfont}
\fi

\long\def\@makecaption#1#2{
   \vskip 10pt 
   \setbox\@tempboxa\hbox{\footnotesize #1: #2}
   \ifdim \wd\@tempboxa >\hsize   
       \leftskip 0pt plus 1fil 
       \rightskip 0pt plus -1fil 
       \parfillskip 0pt plus 2fil 
       \footnotesize #1: #2\par   
     \else                        
       \hbox to\hsize{\hfil\box\@tempboxa\hfil}  
   \fi}

\def\maketitle{\newpage
\null
\vspace*{-2cm}
 {\normalsize \tt \begin{flushright} 
  \begin{tabular}[t]{l} \@date 
  \end{tabular}
 \end{flushright}}
\vspace{2cm}
\begin{center}
   {\bf \@title \par}       
   \vskip 2em                      
   \@aabuffer\relax
\end{center} \par
\gdef\@aabuffer{}
}
\makeatother
\title{
  Top Quark Condensate in Grand Unified Theories\footnote{
Talk presented by M. Tanabashi, to appear in {\it Proc. of
1996 International Workshop on Perspectives of Strong
Coupling Gauge Theories (SCGT 96), Nagoya, Japan, 13--16 Nov. 1996},  
eds. J. Nishimura and K. Yamawaki (World Scientific Pub. Co., Singapore).
}}
\author{
  Iwana Inukai$^{1)}$
  Masaharu Tanabashi$^{2)}$
  and Koichi Yamawaki$^{1)}$
}
\address{
  $^{1)}$ Department of Physics, Nagoya University \\
  Nagoya, 464-01, JAPAN \\ 
  $^{2)}$ Department of Physics, Tohoku University \\
  Sendai, 980-77, JAPAN}
\date{
  TU/97/521 \\
  DPNU-97-29
}
\begin{document}
\maketitle\abstracts{
We 
propose
a top quark condensate scenario embedded in
grand unified theories (GUTs), 
stressing that the gauged Nambu-Jona-Lasinio model has a nontrivial
continuum limit (``renormalizability'')
under certain condition which is actually satisfied in
all sensible GUTs with simple 
group.
The top quark mass prediction in  this scenario is shown to be
insensitive to the ultraviolet cutoff $\Lambda$ thanks to the
``renormalizability''.
We also discuss a possibility to reduce the top mass prediction
in this scenario.
}
\setcounter{section}{-1}
\section{Introduction}

The large top quark mass $m_t\simeq 175$GeV observed by CDF and D0
experiments\cite{kn:cdf97,kn:d097}
is an important clue to explore physics beyond the standard model.
The top quark is coupled to the electroweak symmetry breaking sector
with coupling strength proportional to its large mass.  
It therefore inevitably influences the dynamics of the spontaneous
electroweak symmetry breaking.
Top quark condensate
scenario\cite{kn:mty89,kn:nambu89,kn:marciano89,kn:bhl90} is one of
the most exciting possibilities in this direction.
In this scenario the elementary Higgs field in the standard model is
replaced by newly introduced Nambu-Jona-Lasinio (NJL)-type four-fermion 
interactions between ordinary quarks
and leptons.
Among these four-fermion interactions, one NJL coupling is assumed to
be supercritical, causing dynamical electroweak symmetry breaking.
The quark associated with the supercritical NJL coupling is identified
as the heaviest quark, i.e., the top quark.
This scenario is also referred to as ``Top-Mode'' Standard Model in
contrast to the conventional ``Higgs-Mode'' 
scenario of electroweak
symmetry breaking.

The top condensate scenario suffers from theoretical and
phenomenological difficulties, however.
Theoretical one arises from non-renormalizability of the NJL
interaction\cite{kn:suzuki90,kn:hasenfratz91,kn:justin91}, which
forces us to introduce an ultraviolet cutoff  
$\Lambda$ to regularize loop integrals. 
The cutoff $\Lambda$ is the scale where non-renormalizable NJL
interaction is resolved and new physics takes the place of the NJL
interaction above $\Lambda$.
The precise prediction from the top condensate scenario, therefore,
depends on assumptions made for new physics behind the NJL interaction.
Many references are assuming a sharp momentum cutoff and absence of
higher dimensional interactions other than NJL.
There is no physical reason for these assumptions, however.
Moreover, the predictions from pure NJL model are sensitive to these 
assumption.\cite{kn:suzuki90,kn:hasenfratz91,kn:justin91} 

This scenario also tends to predict too heavy top quark which is
not acceptable phenomenologically.
Since the top quark mass prediction is a decreasing function of the
cutoff $\Lambda$,
we need to assume very large $\Lambda$ to avoid too heavy top quark. 
Even for an extremely large cutoff $\Lambda=10^{19}$GeV, however, the
prediction with the previous assumptions gives too large top quark
mass\cite{kn:bhl90}, $m_t\simeq 220GeV$, still incompatible with the
observed one.

Topcolor\cite{kn:hill91}, which replaces the non-renormalizable NJL
interaction with a
broken topcolor gauge interaction, is a possible candidate to remedy
these difficulties. 
Since the underlying dynamics behind NJL is specified in this model,
there is no conceptual problem arising from the non-renormalizability.
It is also claimed that the topcolor model can avoid serious fine
tuning problem by introducing topcolor dynamics at the electroweak
scale. 
Since the topcolor gauge group can be incorporated into technicolor
model of dynamical electroweak symmetry
breaking (topcolor-assisted
technicolor)\cite{kn:hill95,kn:le95,kn:ae96}, 
the existence of techni-fermion condensate can reduce
the top quark mass prediction significantly.
However, there is no dynamics which can naturally explain why the
scale of the topcolor breaking coincides with the electroweak scale.
The naturalness problem is therefore solved in a rather incomplete
manner in the topcolor model.
Moreover, the topcolor-assisted technicolor would have the notorious
problems of technicolor models, i.e., the large positive
$S$-parameter, $\Delta\rho$ and the excess of the
flavor-changing-neutral-currents.

In this talk, we propose
a new class of top quark condensate scenario in
which the top condensate is embedded in grand unified gauge theories
(GUTs), i.e., top-mode GUTs. \footnote{Preliminary report was given in
  Ref. \cite{kn:Yama96}.}
  The point is that the gauged NJL model (NJL plus
gauge interactions) under certain condition is renormalizable 
in the sense that it has a nontrivial continuum theory in the
$\Lambda\rightarrow \infty$ 
limit.\cite{kn:ksy91,kn:kty93,kn:Yama91,kn:Kras93,kn:hkkn94} 
Such a condition is pointed out\cite{kn:kty93,kn:hkkn94} 
to be equivalent to existence of 
the Pendleton-Ross-type infrared fixed point\cite{kn:pr81}
in the low energy effective (gauged) Yukawa theory, which is actually
realized in all sensible GUTs with simple 
group.\cite{kn:vau82}
Thus we observe that the NJL model becomes ``renormalizable'' when
coupled to GUTs.  
Hence in the class of models we propose, the ambiguities of prediction
caused by the non-renormalizability are washed out by a strongly attractive
infrared fixed point\cite{kn:pr81,kn:vau82} of the renormalization group
equation of GUT-Yukawa coupling. 
We have therefore less ambiguity than the conventional top-mode scenarios. 
We also point out a potential possibility to decrease the top quark mass 
prediction in this model.

\section{``Renormalizability''
of the gauged NJL model}

We here consider the ``renormalizability'' of the gauged NJL model
using the original formulation\cite{kn:mty89} of the top condensate,
i.e., the Schwinger-Dyson gap equation and the Pagels-Stokar formula
for the decay constant of the NG boson.
The renormalization group formulation\cite{kn:bhl90} will 
also be used to discuss the  ``renormalizability''.

Let us 
start with the pure NJL-type interaction in the top-mode
Lagrangian\cite{kn:mty89,kn:bhl90}: 
\begin{equation}
  {\cal L} = \frac{G_t}{2}(\bar q_L t_R) (\bar t_R q_L).
\end{equation}
The gap equation can be written as
\begin{equation}
  m_t = \frac{N_c}{8\pi^2}G_t m_t 
        \left(\Lambda^2 - m_t^2 \ln\frac{\Lambda^2}{m_t^2}\right),
\label{eq:eq2}
\end{equation}
with $\Lambda$ being ultraviolet cutoff.
We obtain a nontrivial solution $m_t\ne 0$ only if the NJL coupling
strength exceeds the critical value:
\begin{equation}
  G_t > G_{\rm crit} = \frac{8\pi^2}{N_c}\frac{1}{\Lambda^2}.
\end{equation}
Although we need a fine-tuning of $G_t$ to obtain finite $m_t$,
the second order phase transition property of
Eq.(\ref{eq:eq2}) enables
us to take such a fine-tuning.
The decay constant of Nambu-Goldstone boson can be evaluated in terms of 
the top quark mass and the ultraviolet cutoff (Pagels-Stokar formula):
\begin{equation}
  v^2 = \frac{N_c}{8\pi^2}m_t^2 \ln\frac{\Lambda^2}{m_t^2}.
\end{equation}
We note here that it is impossible to take $\Lambda\rightarrow \infty$
limit keeping $v$ and $m_t$ finite.
This is 
nothing but 
a consequence of the non-renormalizability of the pure NJL
model. 
In other words the top quark mass $m_t$ vanishes in
$\Lambda\rightarrow\infty$ limit if we keep the decay constant $v$
finite.
This behavior indicates triviality of the effective Yukawa interaction
$m_t/v$.
Prediction of the pure NJL model is therefore very sensitive to
the value of $\Lambda$, the exact definition of the ultraviolet cutoff
$\Lambda$ 
and the existence of higher dimensional operators, i.e., the details of
the physics at the ultraviolet cutoff region.

This situation drastically 
changes if there exists gauge interaction in addition to
the NJL (gauged NJL model): 
The theory becomes ``renormalizable'' in the sense that
it has a finite continuum theory in the limit $\Lambda \rightarrow 
\infty$.\cite{kn:ksy91,kn:kty93,kn:Yama91,kn:Kras93,kn:hkkn94}
To illustrate this phenomenon,
we first consider QCD effects neglecting its running of coupling strength. 
Unlike the case of pure NJL model, the top quark mass as a solution of
the gap equation becomes a function of momentum scale in the gauged NJL
model. 
The asymptotic solution of the gap equation is then given
by\cite{kn:kmy89} 
\begin{equation}
  \Sigma_t(p^2) \simeq m_t 
  \left(\frac{p^2}{m_t^2}\right)^{-(1-\sqrt{1-\alpha/\alpha_c})/2},
\end{equation}
with $m_t$ being the on-shell mass. 
By using this high energy behavior, we may write the decay constant of the
NG boson 
as
\begin{equation}
  v^2 \simeq \frac{N_c}{8\pi^2}\int_{m_t^2}^{\Lambda^2} dp^2 
                               \frac{\Sigma_t^2(p^2)}{p^2}
  = \frac{N_c}{8\pi^2}\frac{m_t^2}{1-\sqrt{1-\alpha/\alpha_c}}
    \left[
       1 -
       \left(\frac{m_t^2}{\Lambda^2}\right)^{1-\sqrt{1-\alpha/\alpha_c}}
    \right].
\end{equation}
Unlike the pure NJL model we can take the $\Lambda\rightarrow \infty$
limit keeping both $v$ and $m_t$ finite and non-zero.
The prediction of the gauged NJL model is therefore insensitive to the
ultraviolet cutoff, i.e., the detail of physics at the cutoff region.
In other words, the gauged NJL model can be
``renormalized'' in the 
sense of its ultraviolet cutoff insensitivity.
We can show the ``renormalizability'' in this sense even in the symmetric
vacuum.\cite{kn:kty93}

This observation, however, might depend on our
over-simplification on the gauge interaction, i.e., the non-running
behavior of the gauge coupling strength.  
We therefore need to study the dynamics of the gauged NJL model
including the running effects.
We parametrize one-loop running of QCD gauge coupling:
\begin{equation}
  \alpha(p^2) = \frac{\pi}{2}\frac{A}{\ln(p^2/\Lambda_{QCD}^2)}, 
  \qquad A = \frac{24}{33-2N_f}.
\end{equation}
Here large $A$ implies slowly running gauge coupling.
The asymptotic solution of the gap equation is given by\cite{kn:my89}
\begin{equation}
  \Sigma_t(p^2) \simeq
  m_t \left(\frac{\alpha(p^2)}{\alpha(m_t^2)}\right)^{A/2}
\end{equation}
which leads to
\begin{equation}
  v^2 \simeq \frac{N_c}{8\pi^2}\int_{m_t^2}^{\Lambda^2} dp^2 
                               \frac{\Sigma_t^2(p^2)}{p^2}
  = \frac{N_c}{16\pi}\frac{m_t^2}{\alpha(m_t^2)}\frac{A}{A-1}
    \left[1-\left(\frac{\alpha(\Lambda^2)}{\alpha(m_t^2)}\right)^{A-1}
    \right].
\label{eq:eq9}
\end{equation}
We note here that we can take the $\Lambda\rightarrow\infty$ limit
keeping both $m_t$ and $v$ finite if $A>1$, i.e., sufficiently slow running
coupling\cite{kn:ksy91}. 
Unlike the analysis within non-running gauge coupling, however,
the cutoff scale physics decouples only through logarithmic suppression.

The dynamics of the gauged NJL model can also be analyzed in terms of 
a low energy induced gauged Yukawa model by imposing a
``compositeness condition''\cite{kn:bhl90} on its renormalization
group equation.
We can further confirm our observation on the 
``renormalizability''
of the 
gauged NJL model by using this technique.\cite{kn:kty93,kn:Yama91,kn:hkkn94}

The one-loop renormalization group equation of the top quark Yukawa
coupling is given by
\begin{equation}
  (4\pi)^2 \mu\frac{d}{d\mu} g_t 
    = (N_c+\frac{3}{2}) g_t^3 - 8 g_t g_3^2, 
  \quad
  (4\pi)^2 \mu\frac{d}{d\mu} g_3 
    = - \frac{8}{A} g_3^3,
\label{eq:eq10}
\end{equation}
with $N_c=3$. 
Here we have neglected effects of the $SU(2)_W$ and $U(1)_Y$
gauge interactions for simplicity.
The compositeness condition, i.e., the absence of the Higgs kinetic
term at $\Lambda$ is described as
\begin{equation}
  1/g_t^2(\mu=\Lambda) = 0.
\label{eq:eq11}
\end{equation}
It is easy to solve Eq.(\ref{eq:eq10}) and Eq.(\ref{eq:eq11}):
\begin{equation}
  g_t^2(\mu) = \frac{8(A-1)}{A(N_c+\frac{3}{2})}
               \frac{g_3^{2A}(\mu)}
                    {g_3^{2(A-1)}(\mu)-g_3^{2(A-1)}(\Lambda)},
\end{equation}
which leads to
\begin{equation}
  v^2 \simeq \frac{2m_t^2}{g_t^2(\mu=m_t)}
   =\frac{N_c+3/2}{16\pi}\frac{m_t^2}{\alpha(m_t^2)}\frac{A}{A-1}
    \left[1-\left(\frac{\alpha(\Lambda^2)}{\alpha(m_t^2)}\right)^{A-1}
    \right].
\label{eq:eq13}
\end{equation}
This expression almost coincides with the expression Eq.(\ref{eq:eq9})
except for a small difference coming from subleading effects in the
large $N_c$ expansion. 

\section{Top quark mass prediction}

In the minimal version of the top condensate scenario\cite{kn:mty89,kn:bhl90}, 
$A$ is given by 
\begin{equation}
  A = \frac{24}{33-2N_f}= \frac{8}{7} > 1, \qquad N_f=6.
\end{equation}
We can therefore naively expect decoupling of the cutoff physics.
Unfortunately this is not true quantitatively in the minimal scenario. 
The decoupling is controlled by the lorarithmic suppression
\begin{equation}
  \left[ \frac{\displaystyle \ln( m_t^2/\Lambda_{QCD}^2 )}
              {\displaystyle \ln( \Lambda^2/\Lambda_{QCD}^2 )}
  \right]^{A-1} \rightarrow 0 \qquad
  \hbox{as}\quad
  \Lambda\rightarrow \infty, 
\end{equation}
which is not enough due to its very slow convergence. 
Actually even if we take $\Lambda\sim M_{pl}$ we obtain
$[\ln(m_t^2/\Lambda_{QCD}^2)/\ln(\Lambda^2/\Lambda_{QCD}^2)]^{A-1}
\simeq 0.8$ due to small $A-1=1/7\ll 1$.

We therefore conclude that 
decoupling of the cutoff physics
due to the ``renormalizability''
of the gauged NJL model, is insufficient in the
minimal scenario.\footnote{
The predictions of the minimal top condensate is very stable for
$\Lambda\simeq 10^{15}\sim 10^{19}GeV$ due to small $A-1$, however.
It can be understood by the ``quasi''-infrared fixed
point\cite{kn:hill81}. } 
The prediction of the minimal top condensate therefore relies on the
assumptions made for the cutoff physics.

\section{``Renormalizability'' of gauged NJL model in GUTs}

The top condensate scenario embedded in a grand unified gauge theory
might be a viable possibility to solve these problems.
In fact, it was shown by Vaughn\cite{kn:vau82} that the condition
$A>1$ is satisfied in all sensible GUTs with simple group, e.g.,
$SU(5)$, $SO(10)$, $E_6$, etc..
{\em Thus the 
``renormalizability''
of the gauged NJL model is naturally realized in GUT scenarios.}

Let us first consider the minimal $SU(5)$ model as an example.
The top quark Yukawa coupling above GUT scale obeys the
renormalization group equation:
\begin{equation}
  (4\pi)^2 \mu\frac{d}{d\mu} g_t 
  = 6 g_t^3 - \frac{108}{5} g_t g_5^2,
\end{equation}
with $g_5$ being $SU(5)$ gauge coupling strength.
The renormalization group equation of $g_5$ is given by
\begin{equation}
  (4\pi)^2 \mu\frac{d}{d\mu} g_5
  = - \frac{108}{5 A} g_5^3
  = -\left(\frac{55}{3} - \frac{5}{6} -\frac{1}{6}N_H 
                        - \frac{4}{3} N_g
     \right) g_5^3,
\end{equation}
with $N_g=3$ and $N_H=1$ being the number of generations and the
number of $\underline{5}$ dimensional Higgs field, respectively.
Here we have assumed $SU(5)$ breaks into the standard model gauge
group by elementary Higgs field of
$\underline{24}$ representation.\footnote{
If we consider dynamical breaking of $SU(5)$, the renormalization group 
equation of $g_5$ can be modified.}

We note here $A= 81/50>1$, suggesting the existence of the infrared fixed
point of Yukawa coupling.
We can therefore take the limit $\Lambda\rightarrow\infty$ with the
``compositeness condition'' keeping finite Yukawa coupling.
We also note that the deviation of $A$ parameter from unity, 
$A-1= 31/50$, is reasonably large, which implies relatively fast
decoupling of the cutoff physics
above the GUT scale.

\section{Top condensate in grand unified theories}
The present measurement of Weinberg angle is not
consistent with the minimal $SU(5)$ GUT\@.
We thus need to introduce non-minimal unification models.
Since we are dealing with the dynamical electroweak symmetry breaking, 
we restrict ourselves within models which do not contain elementary scalar 
particles below the GUT scale.
Particularly, we do not consider SUSY extension of $SU(5)$ model.

Here, we discuss effects of extra fermions in the $SU(5)$ GUT\@.
The minimal extension in this direction was given by Murayama and 
Yanagida\cite{kn:my92}, who
introduced extra vector-like quark $Q$ with $(3,2)_{1/6}$
representation of the standard model gauge group.
The renormalization group equations of the standard model gauge
couplings are modified above the mass of the vector like quark $Q$:
\begin{eqnarray*}
  (4\pi)^2 \mu \frac{d}{d\mu} g_1 
    &=&  -\left(- \frac{1}{10} N_H- \frac{4}{3} N_g - \frac{1}{15} N_Q
          \right) g_1^3, \\
  (4\pi)^2 \mu \frac{d}{d\mu} g_2 
    &=&  -\left( \frac{22}{3}
                - \frac{1}{6} N_H- \frac{4}{3} N_g - N_Q
          \right) g_2^3, \\
  (4\pi)^2 \mu \frac{d}{d\mu} g_3
    &=&  -\left( \frac{33}{3}
                - \frac{4}{3} N_g - \frac{2}{3} N_Q
          \right) g_3^3,
\end{eqnarray*}
with $N_H=1$ and $N_Q=2$ being the numbers of Higgs and extra quark
species, respectively. 
Here $N_Q=2$ implies a pair of extra vector-like quarks.
We can achieve grand unification of gauge 
couplings by taking $M_Q\sim {\cal O}(10^6) GeV$.

The $(3,2)_{1/6}$ representation can be embedded in $\underline{10}$
representation of $SU(5)$.
The renormalization group equation above the GUT scale can be written
as
\begin{equation}
  (4\pi)^2 \mu \frac{d}{d\mu} g_5
    =  -\left( \frac{55}{3}
                - \frac{4}{3} N_g -\frac{5}{6} -\frac{1}{6}N_H - N_Q
          \right) g_5^3.
\end{equation}

It is now straightforward to evaluate the top quark mass prediction in 
the top quark condensate scenario embedded in this particular GUT
model.
We obtain $m_t\simeq224GeV$ for $\Lambda=10^{19}GeV$.
Here we have assumed $\alpha_3(\mu=M_Z)=0.118$.
Unlike the minimal top condensate, we can take
$\Lambda\rightarrow\infty$ limit and obtain
$m_t\simeq 201GeV$ for $\Lambda=\infty$.
The prediction becomes cutoff insensitive thanks to the
``renormalizability'' of the gauged NJL model.

If we allow fine tuning of NJL interaction of bottom quark, 
the low energy effective theory becomes a two-Higgs-doublet model.
In this model, the renormalization group equation of Yukawa coupling
is modified and the prediction of $m_t$ is reduced, $m_t\simeq
193$GeV for $\Lambda=\infty$.
We also note that the prediction of $m_t$ is sensitive to
$\alpha_3(\mu=M_Z$) as indicated in Eq.(\ref{eq:eq13}).
For $\alpha_3(M_Z)=0.110$ we obtain $m_t\simeq 196$GeV, $188$GeV for
$\Lambda=\infty$ in one- and two-Higgs doublet models, respectively.
The $m_b/m_\tau$ ratio can also be calculated in our framework.
We find $m_b\simeq 4.5$GeV for $\alpha_3(M_Z)=0.110$ and
$\Lambda=\infty$ in the  
two-Higgs-doublet model, which agrees with the present measurement of
$m_b$. 

\section{Discussion}

The top quark mass prediction in the previous section was somewhat
heavier than the present measurement.
We therefore discuss how our model can be improved to predict lighter
$m_t$. 

The one-loop renormalization group equation of the effective gauged
Yukawa interaction above the GUT scale can be parametrized by
\begin{equation}
  (4\pi)^2\mu\frac{d}{d\mu}g_t
  = \gamma g_t \left[ B g_t^2 - g_5^2 \right], 
\qquad
  (4\pi)^2\mu\frac{d}{d\mu}g_5
  = -\frac{\gamma}{A} g_5^3.
\end{equation}
The model described in the previous section corresponds to the
set of parameters:
\begin{displaymath}
  A=\frac{81}{50}, \qquad 
  B=\frac{15}{54}, \qquad
 \gamma = \frac{108}{5}.
\end{displaymath}
The solution of the compositeness condition is given by
\begin{displaymath}
  g_t^2(\mu=M_{GUT})
  =\frac{2\pi}{B}\alpha_5(\mu=M_{GUT}) \frac{A-1}{A} 
   \left[ 1 - \left(\frac{\alpha_5(\mu=\Lambda)}{\alpha_5(\mu=M_{GUT})}
              \right)^{A-1}
   \right]^{-1}.
\end{displaymath}
Since we want to construct models which are free from the cutoff
ambiguity, the coefficient $A$ should be large enough.
We therefore obtain
\begin{equation}
  g_t^2(\mu=M_{GUT})
  \simeq \frac{2\pi}{B}\alpha_5(\mu=M_{GUT}).
\label{eq:eq20}
\end{equation}
The GUT scale gauge coupling is constrained by the low energy
measurements of the gauge coupling strength. 
Combining the proton decay constraint for $M_{GUT}$ we obtain the
lower bound of the GUT gauge coupling 
$\alpha^{-1}_5(M_{GUT})<\alpha_{1,{\rm SM}}^{-1}(\mu>7\times
10^{14}{\rm GeV})<40$.
We thus need to construct models with sufficiently large coefficient
$B$  to obtain the top quark mass prediction consistent with the
present measurement. 
Actually such a large $B$ can be implemented in our scenario by a
minor extension. 
The gauge symmetry allows Yukawa interaction between the extra
vector-like quark $Q$, $\underline{5}$ Higgs field and the
$\underline{5}$ fermion\footnote{
Existence of such a Yukawa coupling is required for the
vector-like quark to decay.}. 
If we admit the existence of such a Yukawa coupling, the coefficient
$B$ can be enhanced, leading to a lighter top quark mass, where we
have assumed that the newly introduced Yukawa coupling is proportional 
to the top quark Yukawa coupling.

It is also possible to enhance $B$ parameter by introducing additional 
scalar field and its Yukawa coupling above the GUT scale.
To make the $m_t$ prediction consistent with the present measurement
$m_t\simeq 175$GeV, we need to construct a GUT model with large $B$
parameter, $B\simeq 0.7$.
Here we have assumed Eq.(\ref{eq:eq20}) and renormalization group
equation of the $SU(5)$ GUT containing extra vector-like quarks
below the GUT scale.

We have discussed a top quark condensate scenario embedded in an
$SU(5)$ GUT, stressing the 
``renormalizability'' of the gauged
NJL interaction.
The top quark mass prediction in  this scenario is shown to be rather
insensitive to the ultraviolet cutoff $\Lambda$.
This result can be considered as a consequence of the
``renormalizability'' of the gauged NJL interaction.

This work is supported in part by the Grant-in-Aid of Monbusho 
(the Japanese Ministry of Education, Science, Sports and Culture) 
\# 08640365, \# 09740185 and \# 09246203.

\end{document}